\documentclass[acmsmall]{acmart}
\setcopyright{acmlicensed}
\acmJournal{PACMHCI}
\acmYear{2024} \acmVolume{8} \acmNumber{CSCW2} \acmArticle{488} \acmMonth{11}\acmDOI{10.1145/3687027}

\usepackage{subcaption}
\usepackage{array}
\usepackage{placeins}

\begin{document}

\title{Commit: Online Groups with Participation Commitments}

\author{Lindsay Popowski}
\email{popowski@cs.stanford.edu}
\affiliation{%
  \institution{Stanford University}
  \city{Stanford}
  \state{California}
  \country{USA}
}

\author{Yutong Zhang}
\email{yutongz7@stanford.edu}
\affiliation{%
  \institution{Stanford University}
  \city{Stanford}
  \state{California}
  \country{USA}
}

\author{Michael S. Bernstein}
\email{msb@cs.stanford.edu}
\affiliation{%
  \institution{Stanford University}
  \city{Stanford}
  \state{California}
  \country{USA}
}

\renewcommand{\shortauthors}{Lindsay Popowski, Yutong Zhang, and Michael S. Bernstein}

\begin{abstract}

\begin{figure}[ht]
  \centering
    \includegraphics[width=0.85\textwidth]{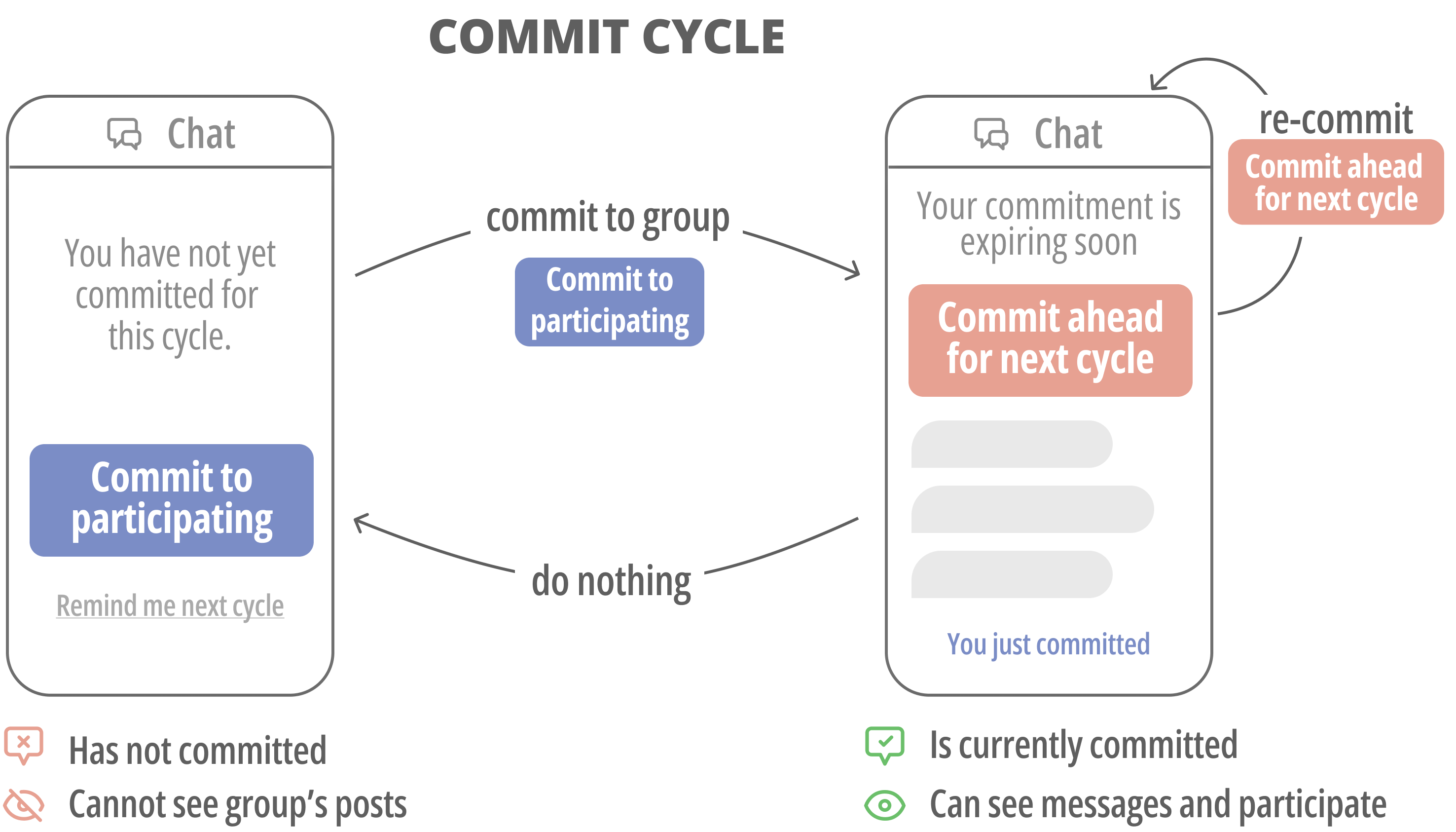}
  \caption{The commitment mechanism and interface: when users are not committed to a group, they do not have the privileges of membership. They must commit to participating in the group in order to join, and renew this commitment on a regular basis to stay.}
  \label{fig: main commit figure}
\end{figure}


In spite of efforts to increase participation, many online groups struggle to survive past the initial days, as members leave and activity atrophies.
We argue that a main assumption of online group design---that groups ask nothing of their members beyond lurking---may be preventing many of these groups from sustaining a critical mass of participation.
In this paper, we explore an alternative \textit{commitment} design for online groups, which requires that all members commit at regular intervals to participating, as a condition of remaining in the group. We instantiate this approach in a mobile group chat platform called Commit, and perform a field study comparing commitment against a control condition of social psychological nudges with $N=57$ participants over three weeks. Commitment doubled the number of contributions versus the control condition, and resulted in 87\% (vs. 19\%) of participants remaining active by the third week. Participants reported that commitment provided safe cover for them to post even when they were nervous. Through this work, we argue that more effortful, not less effortful, membership may support many online groups.

\end{abstract}



\received{January 2024}
\received[revised]{April 2024}
\received[accepted]{Mayy 2024}

\maketitle

\section{Introduction}\label{sec: intro}

Among the defining, multi-decade challenges of CSCW, the challenge of motivating sustained participation in online communities has proven a resiliently difficult one~\cite{beenen2004using, kr_ren2012encouraging, burke_feedme}. Even in modern platforms, forty percent of online groups cease all activity within a week of their creation~\cite{qiu2016lifecycle}, and nearly twenty percent have no activity past the day they are created~\cite{kraut2014role}. With over half of group members never being seen again after their first post~\cite{kr_ren2012encouraging, burke_feedme, yao2021join}, groups fall under the critical mass threshold necessary to sustain activity~\cite{garcia2013friendster,grudin1994groupware,chen2021cold}. Participation is not only necessary for the survival of groups, but also beneficial to the participants: active participants can receive social support~\cite{rau2008intimacy} and the gratification of other social needs~\cite{amichai2016psychological}. Participating in groups is an active form of social media use, which has better mental health outcomes than passive use such as lurking~\cite{verduyn2017social}. The paradox: for many of these groups that fall below the critical mass activity threshold and die out, the same members who are not posting would have preferred that the community survive~\cite{hwang2021why}---suggesting that there is a will, but currently no design interventions to facilitate a way~\cite{ackerman2000intellectual}. 

The most well-known design interventions in research and practice that support group participation follow an assumption that group membership should be \textit{no-commitment}, requiring no ongoing effort to remain a member. This design assumption goes back as far as Usenet: members join with as little as a single click, and may then lurk indefinitely~\cite{nonnecke2000lurker}. 
No-commitment lurking is welcome and even expected for enculturation~\cite{bernstein20114chan}, and it facilitates the kinds of rapid user growth that social computing platforms have historically rallied around, but this laissez-faire approach also creates a bystander-effect-like problem, where members do not feel personal obligation for the fate of the group and do not take action~\cite{martin_diffusion_2015}. For example, 98\% of Reddit users never post or comment 
\cite{reddit_lurker_graphic}. In this regime of no-commitment online groups, our main design toolkit is restricted to nudges. 
Within this paradigm, researchers have developed and studied interventions like goal-setting, gamification, scaffolding contribution, and matching users with particular tasks~\cite{kr_ren2012encouraging}, in the hope that they could sway individuals toward making a Wikipedia edit~\cite{cosley2007suggestbot, lampe2012classroom}, submitting another review~\cite{ling2005using}, or supporting others with similar health problems~\cite{kim2011using, kim2014can}, in the same way that offline volunteer organizations motivate their members. However, despite the existence of these successful interventions for over a decade, issues of limited community survival still persist. For some groups struggling to stay active and alive, this assumption of no-commitment membership may be a poor fit. 

In this paper, we argue that a more successful design direction for these struggling groups may be to make ongoing membership effortful.
We describe a \textit{commitment} design for online groups, which grants a user membership to the group for a limited period of time in exchange for a promise that they will actively contribute to the group, e.g., by posting or replying, at least once during that commitment period. We reinforce commitment as the social contract by making commitments required and visible to other group members~\cite{erikson2000translucence}, creating a clear norm and expectation. When the commitment period ends, the member can renew their commitment by again promising to participate in the next period. Commitment is a familiar way to tie incentives together, as we have to `show up' for many groups in our lives: if we do not join a club's meetings, we are not, in practice, \textit{in} the club; if we do not sit at the lunch table, we do not get to hear what our friends and colleagues are talking about. 
While commitment requirements may not be appropriate for all groups, and there is a risk that some members may leave rather than commit, we argue in this paper and demonstrate empirically that many groups benefit far more than they are hurt by losing their less-committed members.

We instantiate this design in a mobile group chat platform called Commit. Group chats are a popular mechanism for online groups~\cite{nyt_groupchat2024}, and Commit's design mirrors common group messaging platforms such as WhatsApp or WeChat with a few differences to incorporate the mechanism of commitment. First, the group's discussion is not visible until a member makes a commitment to participate in the group. Each member's commitment, made visible in the group's main feed, lasts by default a few days in exchange for the promise to contribute.
Banners and notifications remind the user to fulfill their commitment or recommit for the next cycle when the period is ending and they have not done so. The application does not force participants to fulfill their commitment, nor does it broadcast whether they have done so: we rely on social (dis)approbation as more effective than technical consequences, especially since many groups can survive a little bit of free-riding.

To understand the impact of commitment on online groups, we performed a three week between-subjects field experiment (N=57) where groups of participants were randomized either into a version of Commit as described, or a control condition that mirrored traditional platforms by not requiring commitment. Both conditions contained the same number of notifications and nudges to participants, but different messaging content: the control condition drew on best-practice behavioral nudges to contribute~\cite{kr_ren2012encouraging}, whereas the commitment condition notified users that their commitment period was nearly elapsed without them participating. We examined behavioral data in the form of participation logs, as well as survey and interview data following the study period, to triangulate the overall impact of the commitment mechanism.

Commitment dramatically increased both the levels of contribution and the survival of these online groups. Participants in the Commit condition doubled their median number of messages sent and median number of days active, and experienced five times less attrition. In interviews, participants shared that commitment mechanisms reduced the perceived risk of (re)starting conversations---since everyone had promised to do so---leading to knock-on effects where others replied to the content. Our results demonstrate the promise of orienting online spaces toward communal needs, rather than individual preferences.



In summary, this paper contributes (1) the mechanism of commitment, to encourage more equitable participation in online groups; (2) the app Commit, which instantiates commitment; (3) a field experiment comparing commitment to traditional social psychological nudges; and (4) a description of the design space for commitment, which includes advice for different group types, as well as possible extensions of our intervention method and recommendations for future efforts towards thriving online groups.

\section{Related Work}\label{sec: related work}

While significant effort has been expended to understand the causes of online group decline and mitigate it through encouraging participation, this problem persists. We are interested in drawing from a wide swathe of psychological and sociological insights regarding how group members and their behavior influence one another, and how offline groups engage with that influence. This work contributes to a broad history on encouraging participation and contribution online, as well as a more concentrated set of work that engages with this goal through new affordances and systems. 

\subsection{Background}
The question of commitment and contributions to online groups has a long history at CSCW. We begin by reviewing this central framing literature, then turn to related work that specifically motivates and frames our approach of introducing commitment mechanisms.

\subsubsection{Online Participation and Group Survival}

Overwhelmingly, online groups do not survive. On Facebook, 19\% of groups had no activity after the day they were created, and fewer than half were still active after three months~\cite{kraut2014role}. Similarly, around 40\% of threads on Usenet~\cite{joyce2006predicting, burke2007introductions} and 43\% of threads on 4chan's /b/ discussion board~\cite{bernstein20114chan} never receive even a single reply. Among WeChat groups, out of several million new groups a day, around a third die within their first few days~\cite{qiu2016lifecycle}. Group death is not just a signal that users are turning to other groups: members are highly likely to become inactive on the entire platform in their early days. Around two thirds of newcomers to Usenet post once and then never again~\cite{krautresnicknewcomers}, similar to the 62\% of health support forum members who never return after their first day~\cite{yao2021join}. MMORPG guild members and WeChat users exhibit similarly severe attrition in their groups, even if they remain on the platform~\cite{krautresnicknewcomers, zhang2016comeandgo}.

The early days of online groups are fraught, both for the group as a whole and for its members. One framing sources this problem back to insufficient membership. Group survival is then an effort to overcome the cold start or critical mass problem, where groups struggle at their conception to reach enough users to be self-sustaining~\cite{grudin1994groupware, chen2021cold}. Attracting members is difficult because online communities can provide value only when there are multiple active participants, given that they are primarily instruments of connection and communication. Sociotechnical systems often have paradox of needing participants in order to produce value but needing that value to draw and keep participants, and this framing encourages a path of rapid initial growth to persevere.

However, not all communities want to be enormous in order to survive. In many small groups, the size offers desirable traits: specificity of purpose, less toxicity, and higher quality content, leading members to prefer them to remain that size~\cite{hwang2021why}. In addition, group growth may oppose other goals that groups aspire toward, like community cohesion~\cite{hwang2021why} and member retention~\cite{butler2001membership}. Diffusion of responsibility means that growth can negatively impact the effort that members put into the group, since the responsibility for responding, helping others, organizing, and other tasks decreases for each individual member \cite{martin_diffusion_2015}. We therefore aim to support group survival by focusing on the existing members, not adding new ones~\cite{cunha2019alike}. While new members are unlikely to post~\cite{burke_feedme}, in the group's early days there is no one else to sustain the group. Since member activity is a strong factor influencing community survival versus collapse~\cite{garcia2013friendster, kraut2014role}, we turn to encouraging participation from group members to help groups survive.

\subsubsection{Encouraging Participation Online}

For close to as long as online communities have existed, research has addressed the question of how to encourage contribution. Attempts to encourage contribution often involve (a) strategically asking users, (b) gamifying contribution or creating new rewards, and (c) appealing to or enhancing intrinsic rewards~\cite{krautresnickcontributions}. For example, one approach motivated users to rate more movies using techniques drawn from social psychology, like highlighting uniqueness by informing users they had insights that others would not be able to contribute~\cite{ling2005using, beenen2004using}. These efforts exist across many contexts: increasing Wikipedia edits via routing the right tasks~\cite{cosley2007suggestbot} or including it in their education~\cite{lampe2012classroom}, encouraging more social support in health communities via adding membership statuses~\cite{kim2011using}, adding reactions to scaffold participation in learning~\cite{nacu2015underrepresented}, and encouraging forum participation in massive open online courses~\cite{kizilcec2014encouraging}.

These approaches have largely abided by existing default designs for online groups, which specify that members ought to remain members unless they explicitly leave the group. Lurking is considered a default. In this work, we explore more direct changes to this social contract in the form of commitment, requiring that members take effortful action to remain in the group environment. While prior approaches are attractive because they work within existing frameworks and platforms, we explore commitment explicitly as an alternative that proposes deeper changes to the social contract. 

\subsubsection{Understanding Groups}

Members' participation creates social rewards that can motivate others to contribute. Prior work suggests a causal model of how individual and group outcomes interrelate with individual motivation: a member's effort leads to both an individual and a group outcome (along with other members), which together provide utility that motivates the member~\cite{karau2014collectiveeffort}. This motivation structure causes problems like social loafing to emerge, in which individuals have reduced motivation and expend less effort in group settings than when alone~\cite{karau1993social}. However, through the K{\"o}hler motivation gain effect, group members can put in higher effort if they are meant to feel indispensable---uniquely needed---and identifiable---visible in their contribution or lack thereof~\cite{kerr2011kohler}. Goalsetting in groups can help supplement those two factors to improve group outcomes~\cite{thurmer2017planning}. 

We look also to offline parallels to enhance our understanding of the forces dissuading participation. Social movements are a salient point of reference, due to the strong cost and benefit factors at play. Like our online group context, potential participants in a social movement have limited information about whether others will participate and have to make their own choice before they have certainty~\cite{klandermans1984mobilization}. They instead analyze the likely risks and benefits, through estimating factors like the likely impact of their contribution as well as the actions and effects of others~\cite{klandermans1984mobilization}. This knowledge suggests that we can increase participation by creating greater certainty about risks and benefits, which has been successful in the realm of collective action~\cite{cheng2014catalyst}. 
Our approach for commitment acknowledges this need for reassurance and combines it with strategies of goal-setting and group and individual motivation to encourage contribution and help groups survive. 

\subsection{Contribution Expectations and Evolution}

Researchers have urged a more expansive understanding of group members beyond the ``contributors'' versus ``lurkers'' dichotomy~\cite{edelmann2017lurking, takahashi2007active}, one which the commitment mechanism leans into. Groups have different needs and expectations in terms of member commitment, and lurkers could be desirable participants for some. In spaces like LiveJournal, they are thought of as members of the community rather than peripheral layabouts~\cite{kate2012lurking}. Commit does not oppose these communities---lurker-driven communities are valuable to the web---rather, we emphasize that it is important to  open the door to other types of groups as well. The status quo design of most online platforms supports lurker-friendly communities, while Commit seeks to meet the needs of communities which are struggling to survive under this paradigm. Some groups prefer smallness~\cite{hwang2021why}, which may not be compatible with high lurker populations, due to the critical mass of activity needed to sustain a group. However, if community leaders seek to adapt Commit to a lurker-friendly environment, the concept of legitimate peripheral participation~\cite{lave2001legitimate} can inform our effort to build low-effort and low-visibility commitment options.

Commit was designed with membership lifecycles---where members contribute in different ways as they join, meet their own needs, and then fade out~\cite{yao2021join}---in mind, and therefore makes recommiting the active choice so as to make fading out natural and less visible. The needs of groups and their members may change over time, especially as members build greater comfort in the group and take on larger responsibility. Members could be scaffolded through different roles using their type of commitment, inspired by the reader-to-leader framework~\cite{preece2009reader}.

\subsection{Systems to Encourage Contribution and Improve Virtual Interactions}

Previous systems have sought to create new automated ways to encourage contribution. This can be done with automated recommendations: SuggestBot used intelligent task routing to deliver appropriate tasks to contributors with the aim of increasing Wikipedia edits~\cite{cosley2007suggestbot}. Systems can also uncover the best ways to encourage interaction, such as a socially situated question-asking bot which learns which questions people are inclined to answer through live interaction with social media users~\cite{krishna2022socially}. Other systems attempt to incite more interactions by lowering the threshold of interaction and developing new interaction modalities. For instance, Nooks and the Notification Collage both aim to create new instances of workplace socialization through generating opportunities for people to stumble upon relevant topics and discussions~\cite{bali2023nooks, greenberg2001collage}. Systems may also develop new modalities for low-effort sharing to scaffold towards greater intimacy~\cite{liu2021otter, greenberg2001collage}.

Social computing systems have also sought to help online groups collaborate effectively. Some systems focus on artistic collaboration, allowing groups to create together through scaffolding and organizing the tasks involved~\cite{kim2014ensemble, cong2021collective}. These artistic collaborations may also provide social and emotional benefits through building mutual understanding among the participants~\cite{cong2021collective}. Other collaboration-oriented systems focus on helping members or leaders handle the scale of the group through synthesis~\cite{zhang2018making} or steering feedback towards more aggregable forms~\cite{jasim2021communityclick}. 

A few systems have implemented commitment requirements as gates to \textit{group formation}, in which sufficient participants must commit to contributing before a group can form, in the social movement~\cite{cheng2014catalyst}, information-sharing and support~\cite{area51_2024}, and workplace socialization contexts~\cite{bali2023nooks}. Commit speaks most directly to these systems, though it sets limits on individual membership, rather than group existence.

\section{Commitment-Based Group Design} \label{sec: system}

Current designs offer communities insufficient support to start out and survive. Online groups struggle to reach a sustaining level of participation and to maintain it to survive, and the current paradigms to encourage participation, while helpful, are not enough to support the group in its goals. There is a promising opportunity to adapt the framework of online group membership to more clearly align individual incentives with actions that support group survival and thriving. We first explain the design guidelines that brought us to the idea of commitment. Then, to explore this potential space of effortful membership, we describe the design of the mechanism of \textit{commitment} and instantiate it in the group messaging app called \textit{Commit}. 

\subsection{Design Guidelines}

Our high-level motivation: for some groups, having fewer but highly active members might be a better outcome than having more but oft-lurking members. While there are many possible means to achieve this goal, we pursued one focused on raising the expectations for membership. How do we design such a mechanism in a way that enhances the group, rather than strangles it? 

One major cause of lurking is the perception among members that their contributions are not expected or needed in order to achieve good group outcomes~\cite{amichai2016psychological, grudin1994groupware}. Another significant factor is the member's uncertainty or low expectations of gratification from posting, estimated from the quality of responses they have received and the reception of other members when they contribute~\cite{amichai2016psychological}.

\setlength{\tabcolsep}{6pt}
\renewcommand{\arraystretch}{1.5}

\begin{table}[tb]
\begin{center}
\small
\begin{tabular}{p{1.4in} p{1.75in} p{1.75in}} \\
\hline 
\textbf{Characteristic} & \textbf{Status Quo Group} & \textbf{Commit Group} \\
 \hline
 \textbf{Expectations for \newline participation} & No expectation for participation & Expected to send at least one message every two days \\
 \textbf{Perspective towards \newline participation} & Participation as self-serving & Participation as benefiting the group and one's self \\
 \textbf{Expectations of others} & No expectations of others & Expectation of support and \newline reciprocation from others \\
  \textbf{Source of effort} & Either self or others put in effort & The group puts in effort as a team \\
 \textbf{Model of membership} & Membership lingers past \newline presence & Membership is indicative of presence \\
 \hline 
\end{tabular}
\end{center}
\medskip
\caption{Through commitment, aspects of the relationship between member and group are changed.}
\label{table: bitflip}
\end{table}

In response, we chose commitment as a central design lever and formulated two central design guidelines. First: we need to \textit{explicitly align individual incentives with group goals}. Since people frequently do not take personal responsibility for contributing or consider how their individual behavior affects the group, we should \textit{make the group’s expectations for participation explicit}, to communicate the need and institute a norm. Since norms are more effective when made salient to ongoing decisions~\cite{cialdini1991focus}, the system design should also \textit{privately remind users when they are not meeting the group expectations}. Second: \textit{commitments should be explicit and visible to group members}~\cite{erikson2000translucence}. Critical mass problems emerge because participants or potential users do not want to invest effort for uncertain benefit~\cite{grudin1994groupware} and may occur within a social context if individual members have low confidence in the responses or future activity level of other group members. So, commitment-based systems should explicitly \textit{formalize the intention of group members to participate, require this explicit agreement from each member, and make it visible to the group} so as to provide confidence that everyone who is present will be contributing.

In response to these design guidelines, we designed a simple, recurring group commitment mechanism. Groups set particular expectations for participation on a periodic basis, for example posting once a week. In order to be in the group during a particular cycle and see the group content during that time, an individual is required to commit to participating during that cycle. The commitment cycle is synchronized across the group, starting and ending at the same time for every member to draw them into the group at the same time.

\begin{figure}[tb!]
  \centering
    \includegraphics[width=0.9\textwidth]{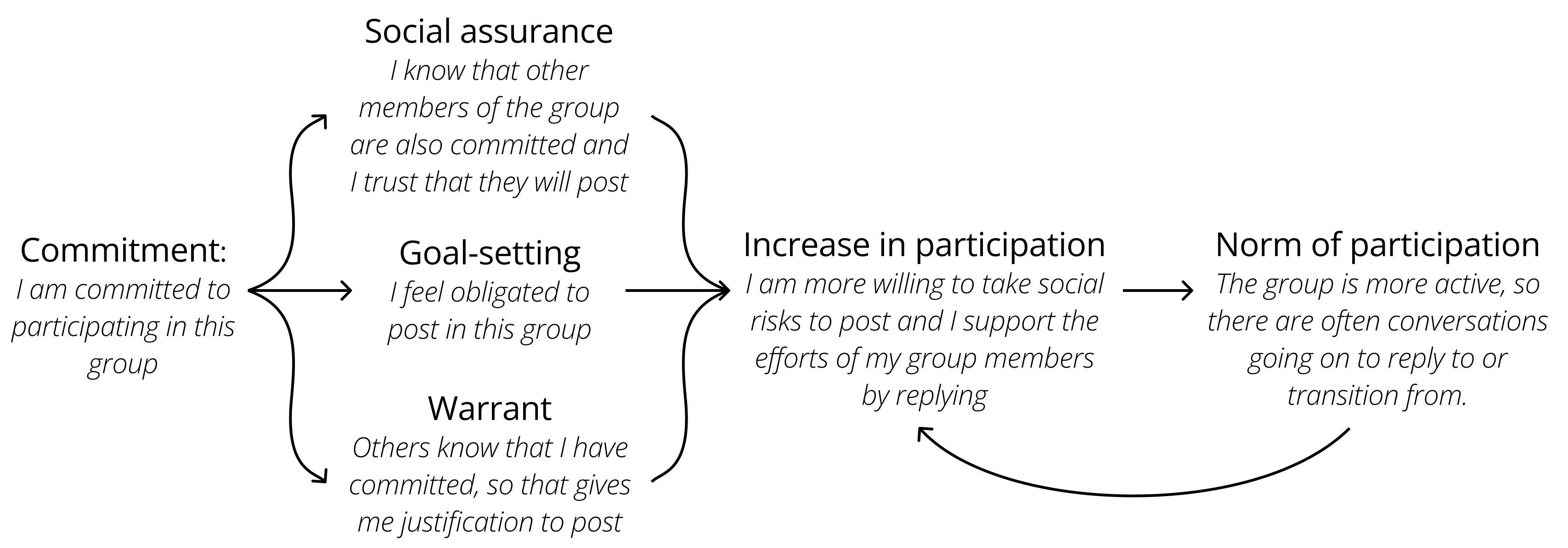}
  \caption{A description of commitment's main effects. Committing assures participant of the intentions of other group members, sets an achievable participation goal for them, and provides a warrant for taking initiative in the conversation. Together, these factors lead to an increase in participation from group members, which over time creates a norm of participation which further encourages participation.}
  \label{fig: commit mechanism}
\end{figure}

Committing is an explicit action, in which users click a button to promise the rest of the group that they will participate during the current cycle. As they do so, they gain access to the group during that time. Commitments display in the feed to the rest of the group (see Figures~\ref{subfig: commit, no ahead, no fulfil}, \ref{subfig: commit, no ahead, fulfil}). The system reminds users of two things: when their commitment is expiring, so that they can commit ahead for the next cycle; and when the cycle is reaching its end and they have not yet participated at the expected level, so that they can fulfill their commitment. While an application could easily block members from seeing the group if they don't fulfill their commitment, or publicize that users fell through on their commitment, we choose not to pursue this kind of public shaming or automatic punishment. Instead, since commitments are regularly broadcast to the rest of the group, we rely on group members to recognize and back-channel to members who are regularly not contributing, as well as the possibility of such a confrontation dissuading social loafing in the first place.

Commitment intends to shift the social contract between members and their group in several ways (Table~\ref{table: bitflip}). Through making membership effortful, commitment creates expectations of participation for members and teaches members to think of their participation as serving the group, and not just themselves. Users also expect their fellow group members to reciprocate their effort and support their endeavors, to support the group as a team. Commitment also rethinks membership to be a dynamic label that indicates active presence and can fluctuate with someone's current capacity. These factors produce multiple motivating processes---social reassurance, setting goals, and providing a warrant---that together encourage participation in group members (Figure~\ref{fig: commit mechanism}).

\subsection{Commit group chat application}

Commitment-based designs can be adapted to different existing platforms that facilitate online groups or communities: Slack, Discord, Reddit, Facebook, and WhatsApp, among others. We created a stand-alone mobile app to realize the approach so that commitment could be a first-class element of the design rather than a side integration. In our group chat application, which we call Commit, the commitment framework is implemented within a simple group chat interface, reminiscent of GroupMe, WhatsApp, Facebook Messenger, or iMessage. 

\begin{figure}[tb]
  \centering
  \begin{subfigure}[t]{0.31\textwidth}
    \centering
    \includegraphics[width=\textwidth]{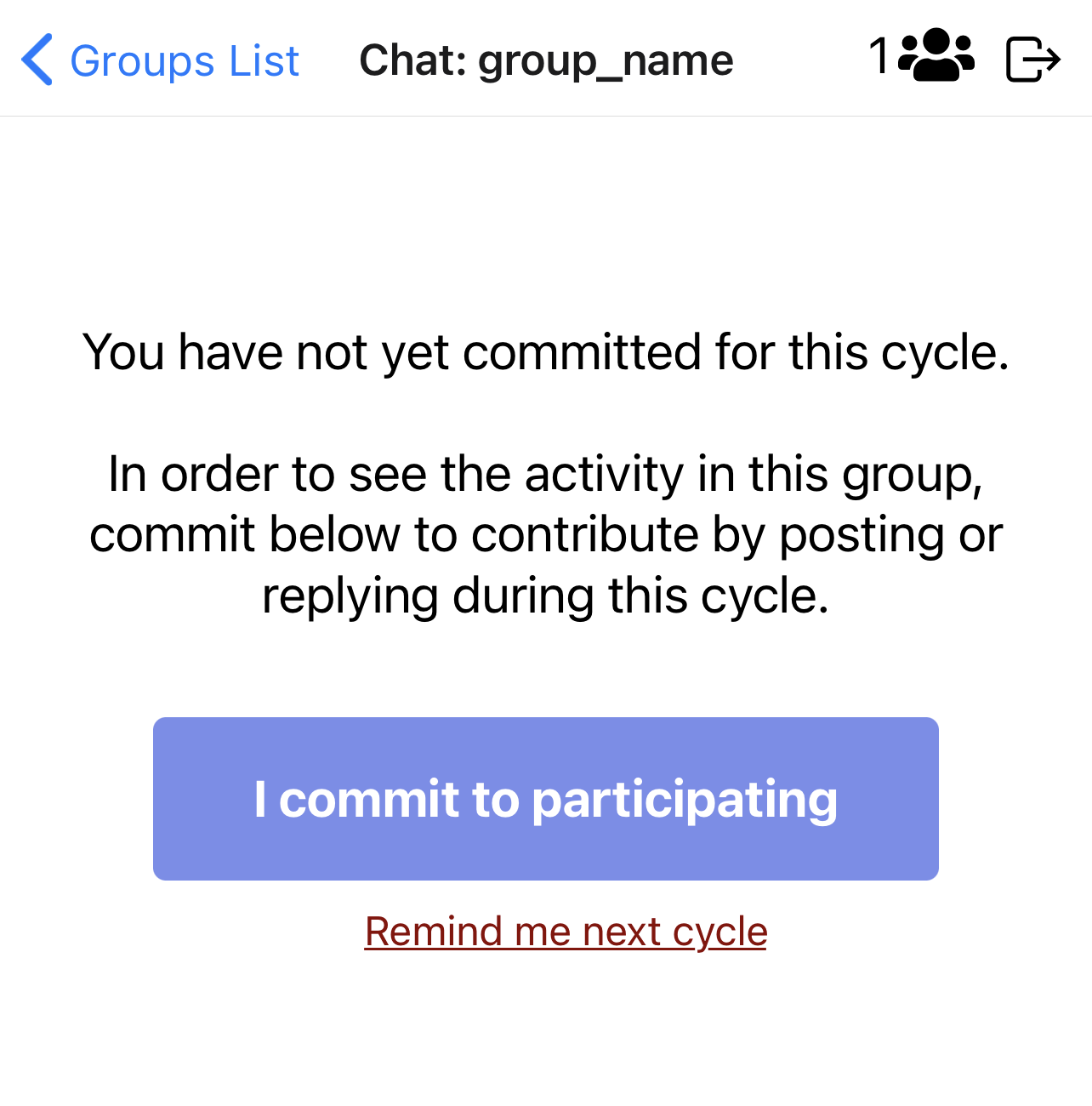}
    \caption{Before committing for the first time or after commitment has lapsed. User must commit to access group messages.}
    \label{subfig: no commit}
  \end{subfigure}
  \hspace{0.01\textwidth}
  \begin{subfigure}[t]{0.31\textwidth}
    \centering
    \includegraphics[width=\textwidth]{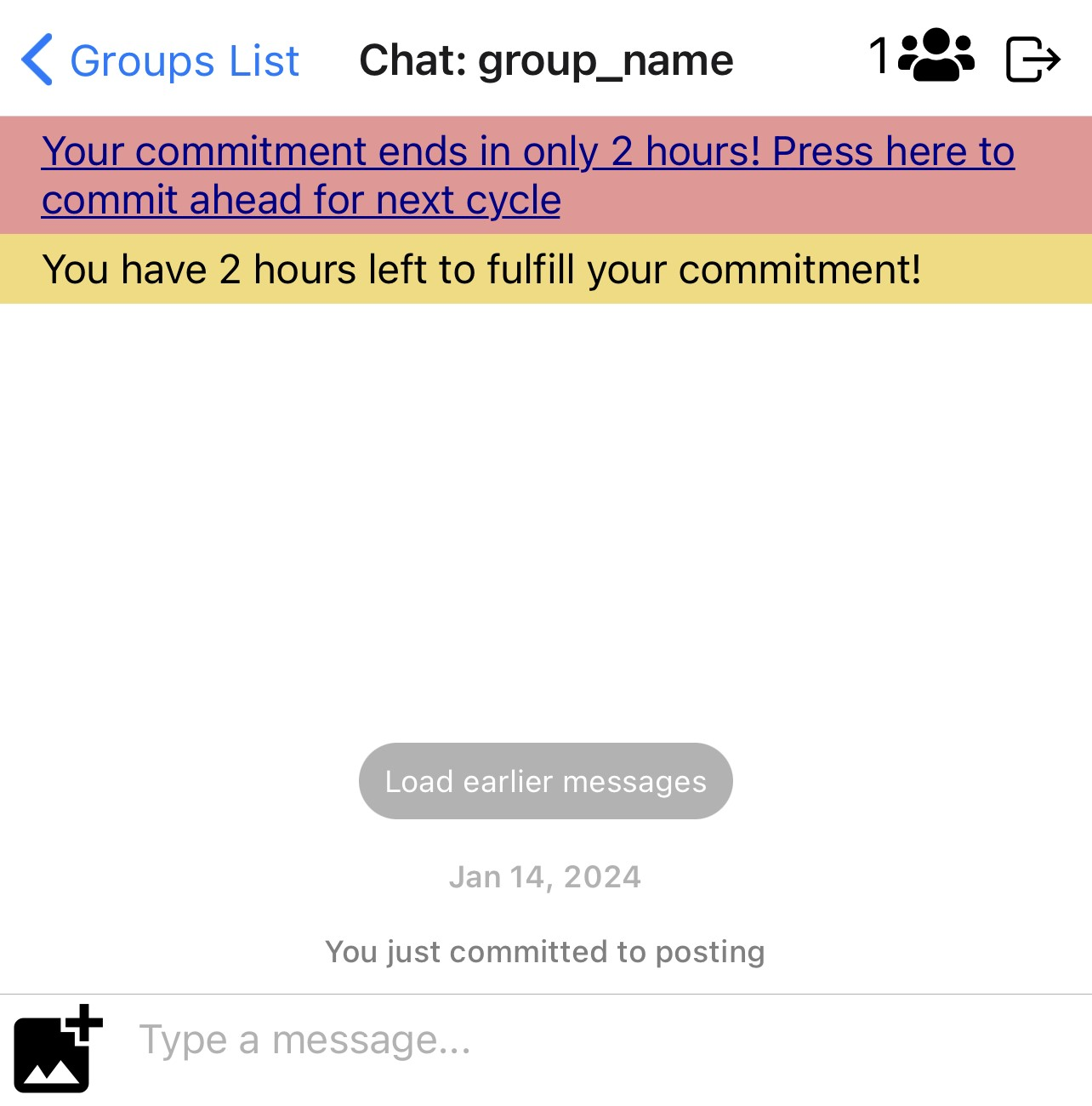}
\caption{Committed for the current cycle but not for the upcoming one. Has not fulfilled the current commitment.}
    \label{subfig: commit, no ahead, no fulfil}
  \end{subfigure}
  \hspace{0.01\textwidth}
  \begin{subfigure}[t]{0.31\textwidth}
    \centering
        \includegraphics[width=\textwidth]{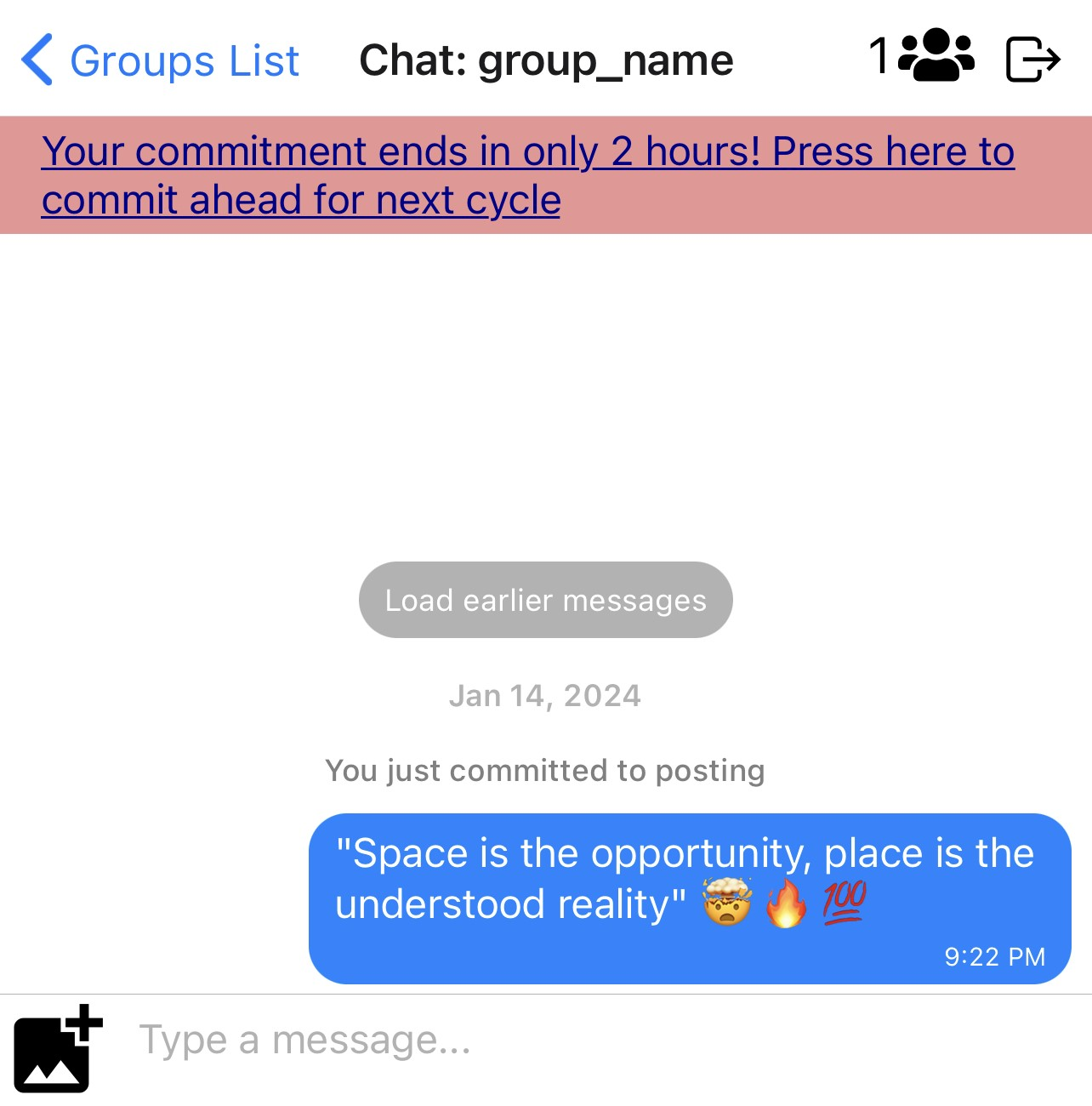}
    \caption{Committed for the current cycle but not for the upcoming one. Has fulfilled the current commitment.}
    \label{subfig: commit, no ahead, fulfil}
  \end{subfigure} 
  \caption{Commitment cycle interface in Commit. A user first comes to a group (a), commits to it and can see the group (b), and then sends a message thereby fulfilling their commitment (c).}
  \label{fig: app prototype}
\end{figure}

In Commit, groups operate on a commitment cycle of every two days, and the participation expectation is to send a message once during each cycle. Only committed members are able to see the chat; otherwise, they see an obscured view of the group and a promise that they can see the content once they commit (Figure~\ref{subfig: no commit}). Both committed and non-committed users can see the group name and the number of committed members. When committed members view the group chat, a banner at the top of the channel privately reminds users when their commitment is expiring or if they have yet to fulfill their commitment (Figure~\ref{subfig: commit, no ahead, no fulfil},~\ref{subfig: commit, no ahead, fulfil}). The banner changes colors to indicate urgency when the commitment cycle is close to ending. A separate screen shows the users currently in the group and the last time they posted in the group. Users are not punished for letting their commitment lapse: they can always commit partway through the cycle without penalty, and see all the content they missed. They may also commit early if they want to avoid lapsing: for example, a user may commit one cycle in advance, ensuring that they don't miss anything. 

Notifications are a necessary part of group chats, to keep members aware of new posts. We use mobile notifications for (a)~reminders and (b)~actions of group members. Specific reminders are sent out when the user's commitment will lapse soon, when the user's commitment lapses, once after a long period of absence (several commitment cycles), and if the user's commitment period is ending soon without them having fulfilled their commitment. As in a typical messaging application, users receive notifications when other members send messages, or when someone reacts to their message. If a user is not committed, they still receive notifications when a member posts, but they cannot see the content of what that member posted. Users can manage notifications through their mobile phone settings.

Otherwise, the Commit application replicates the main affordances of a group discussion platform. Users can send text and image messages, and have a small selection of emoji reactions. There is also a special commitment reaction, allowing users to explicitly re-commit in reaction to another member’s message. This action permits a more weighty show of appreciation than a ``like'' reaction, because it is tied to a specific promise of contributing to the group.

\section{Evaluation: Method} \label{sec: eval}

We designed Commit to increase activity in online groups and prevent them from deteriorating into ghost towns. In this section, we investigate through a field experiment whether this approach does in fact increase contributions to, and survival of, online groups. We operationalized this question into three central hypotheses that test its behavioral effects on activity and its attitudinal effects on psychological safety:
\begin{itemize}
    \item \textit{Hypothesis 1}: Participants will send more messages when using Commit.
    \item \textit{Hypothesis 2}: Participants will stay active for more days when using Commit.
    \item \textit{Hypothesis 3}: Participants using the Commit condition will experience greater psychological safety in the group.
\end{itemize}

To test these hypotheses, we performed a three week between-subjects field study, in which participants were assigned to groups of 4-5 members, and groups were randomly assigned to either a control condition or a Commit condition. From this study, we analyze both behavioral data---logged actions such as commits, posts, and reactions---and attitudinal data---semi-structured interviews and surveys with the participants---to understand the social and individual impact of the Commit system. We triangulate the impact of Commit through these three data sources.

\subsection{Study Operations}

\subsubsection{Participants}

We recruited participants for a three-week study on group messaging from Slack workspaces at our institution and email lists, as well as an internal course research pool at our university. We also asked that people forward the recruitment message, so our participant population includes those at other institutions and outside academia. We matched them into groups of 5 people, though some groups ended up with 4 members if one of the matched members never joined the group. Participants were compensated with either a \$50 Amazon gift card or course extra credit (if they signed up via a course). In total, we had 57 participants across 12 groups, with 27 participants (6 groups) in the control condition and 30 participants (6 groups) in the Commit condition. 

\subsubsection{Procedure}

When participants signed up for the study, they indicated topics that they found interesting and, optionally, participants to be grouped with. They were then matched into groups of 5 participants with similar interests. While many participants shared an ingroup with other members (e.g., students at the same university), most did not directly know each other prior to joining a group together. Participants downloaded the Commit app to their iOS or Android phones and enabled notifications. They then joined their assigned group and sent a message introducing themselves to the rest of the group. During the three-week study period, we asked participants to make an effort to chat in the group, but specifically noted that they did not have to message in the groups at a particular rate (both conditions) or commit any specific number of times during the study period (Commit condition). At the end of the group chat use period, participants completed a survey and a semi-structured interview about the atmosphere and behavior of the group.

\subsubsection{Control Condition}

Both conditions used the same custom group chat app that we developed, but the control groups interacted with a version that disabled the commitment mechanism. Their in-chat banner was a reminder of how long it has been since they last posted. They received notifications at the same rate as the Commit groups. Instead of notifications around commitment, participants in the control condition received notifications reflecting best practices in encouraging participation, like nudging, reminding people of intrinsic benefits, and requesting specific desired actions~\cite{kr_ren2012encouraging}. We show an example of paired notifications in Figure~\ref{fig: notifications}; full notification text is in Appendix~\ref{appendix: notifications}. 
\begin{figure}[htb]
  \centering
    \includegraphics[width=0.7\textwidth]{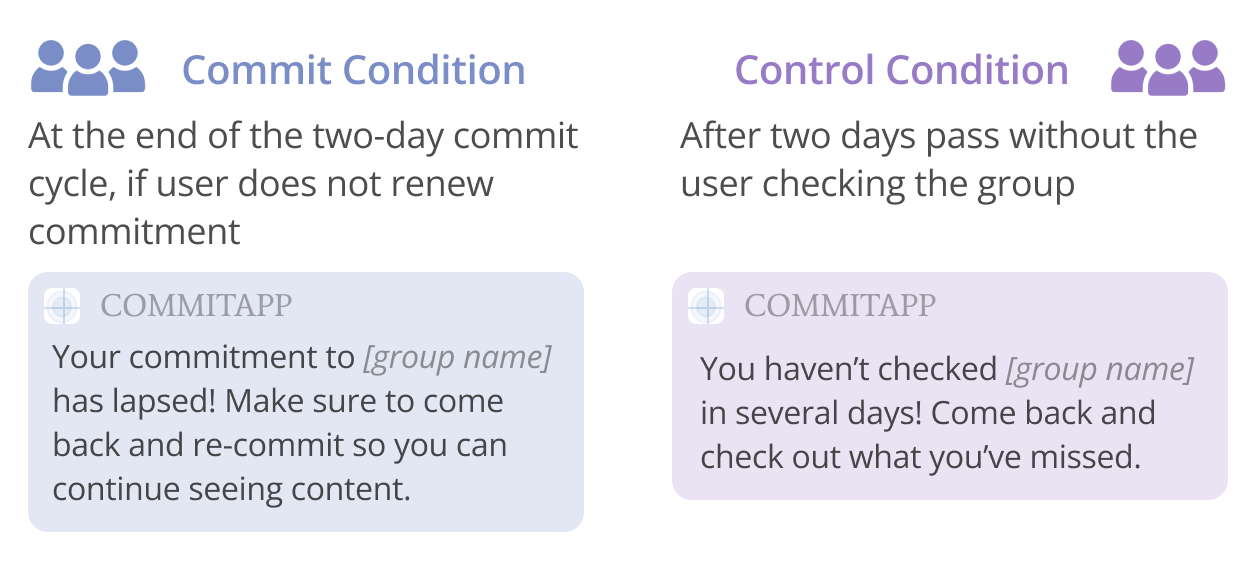}
  \caption{Participants in both conditions received push notifications to draw them back to the app and remind them to contribute. We paired notification types across conditions so that participants received them on a similar cadence and were encouraged to take similar actions; the framing was what changed.}
  \label{fig: notifications}
\end{figure}

\subsection{Measures and Analysis}

\subsubsection{Behavioral Measures}
We focus our analysis on three activity measures of interest for group members: overall messages sent, activity (whether a member posted to the group on a given day), and length of survival. In our analysis of total messages sent by individual participants and groups, we follow the example of prior work~\cite{burke_feedme} in using the logarithm of the number of messages per user to control for skew. We analyze our results using a linear mixed effects model, accounting for the random effects of participant. To analyze activity as a binary state of ``active'' or ``inactive'', we define an individual to be active on days when they send at least one message. For this analysis, we used a mixed effects logistic regression. To consider the survival of individual activity, we define an individual as surviving up until they spend a total of one week inactive, even if they regain activity afterwards.\footnote{We analyze these results for 3, 5, 7, 9, and 11 days to check for robustness. For settings of 3-9 days, the number of days does not impact the directionality or significance of our results.} We use a Cox proportional hazards regression model to compare survival rates.

\subsubsection{Survey Measures}

We administered short surveys to every participant at the end of the study period. Three participants completed the group chat part of the study but did not complete the survey. Questions focus on four themes of interest: feeling valued and important, psychological safety, commitment to the group, and participation inequality. Each theme contains three questions, answered according to a seven-point Likert scale. We include the survey questions in Appendix~\ref{appendix: survey}. 

\subsubsection{Semi-Structured Interviews}

We conducted semi-structured interviews towards two goals: (1) to uncover the subjective experiences of the participants in the different conditions of the group chat, and (2) to understand possible mechanisms that explain our results. The main question topics involved group atmosphere, activity level, equality of participation, and their motivation to participate or not (more detail on questions found in Appendix~\ref{appendix: interview}). All but five participants completed the interview. The first and second authors conducted the interviews over the Zoom videoconferencing software, recording and automatically transcribing them, and then corrected the transcripts by hand. 

We used a thematic analysis approach to reveal patterns across our qualitative interview data~\cite{braun2006using}. First, we performed a line-by-line coding of the transcripts; some of the codes were inductive, emerging directly from the data, while others were informed by our guiding questions. The first and second authors each coded part of the transcripts, meeting to discuss and refine codes. Once all of the transcripts had been coded using the solidified codebook, we met several times to synthesize our line-level codes into larger themes. Many of these themes related strongly to our interview questions, though others were emergent. Once these themes had largely settled, we searched within and between themes for patterns of meaning, also making comparisons across conditions to understand the differing experience of those using the Commit system versus the control. 
\section{Evaluation: Results} \label{results}
In this section, we triangulate across our three data sources---activity logs, survey data, and interview responses---to understand the impact of our Commit intervention. 

\subsection{Commitment doubles levels of participation}

\begin{figure}[htb]
  \centering
  \begin{subfigure}{0.40\textwidth}
    \centering
    \includegraphics[width=\textwidth]{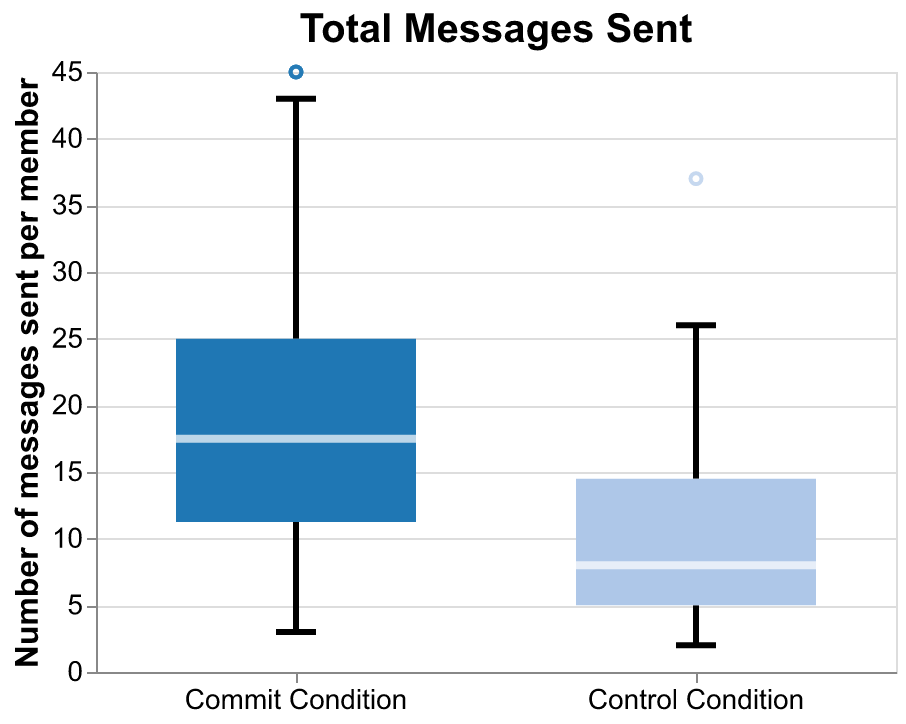}
  \end{subfigure}
  \hspace{0.02\textwidth}
  \begin{subfigure}{0.40\textwidth}
    \centering
        \includegraphics[width=\textwidth]{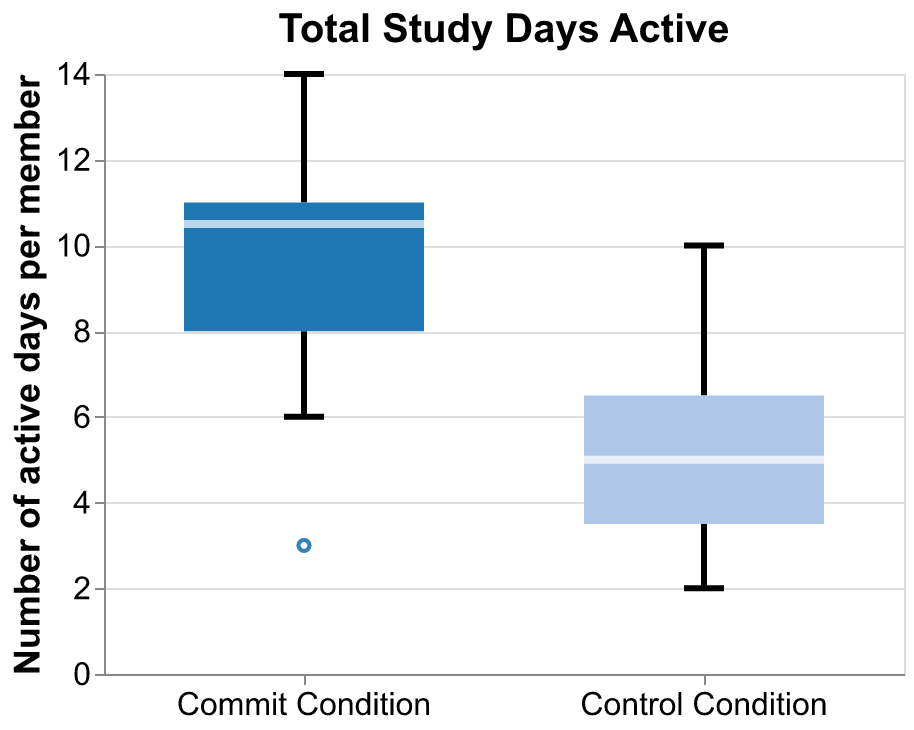}
  \end{subfigure}
  \caption{Comparisons of member participation in groups based on two factors: (a) the number of messages sent by group members during the study period, and (b) the active status of members in the group on each day of the study. Members in the commitment condition sent more messages (around twice the median number) and were active on far more days (over two times as many, according to the median)}
  \label{fig: participation boxplots}
\end{figure}

The Commit condition strongly increased participation rates across multiple measures (Figure~\ref{fig: participation boxplots}), supporting Hypothesis 1. Participants in the Commit condition sent a median of 17.5 messages and were active a median of 10.5 out of 21 days, in contrast to 8 messages and 5 days for the control condition (a factor of 2.2 for messages, and 2.1 for days active). The mixed effects logistic regression for activity and linear mixed model for number of messages (Table~\ref{table: activity model results} in Appendix~\ref{appendix: activity}) both report significant coefficients for condition at the $p<.01$ and $p<.001$ levels, with Commit featuring an odds ratio of 3.5 (log odds: 1.257) on daily activity and an increase in the logged number of messages by 1.143 (an increase by roughly a factor of 3).

Participants were generally happier the more active the group was. Commit groups could be much more active than participants expected: 
\begin{quote}
    \textit{I think I was surprised, and I appreciated everyone's willingness to just communicate. I think I was expecting maybe like-- I think I was expecting more hesitance to be open to having conversations with like random people you don't know.} (P49)
\end{quote}
Despite the possibility that the Commit condition might lead to perfunctory contributions (e.g., ``I'm posting to fulfill my commitment''), participants reported in interviews that most messages felt effortful and none that degree of perfunctory. 
Some participants in the Commit condition admitted to sending less effortful messages in response to their notification prompts, but these messages were still relevant. For example, P23 shared that when choosing to start a new conversation, ``I'll say something like, `Oh, like, what's everyone's weekend plans?'''

In terms of how members perceived the activity level, most Commit participants described it as somewhat regular check-in style conversations, with the occasional more engaged conversation. In contrast, participants in the control condition were likely to describe a few conversation attempts, some of which died out and some of which caught on and led to back-and-forth discussion.

\subsection{Commitment increases group survival}

\begin{figure}[htb]
  \centering
    \includegraphics[width=0.7\textwidth]{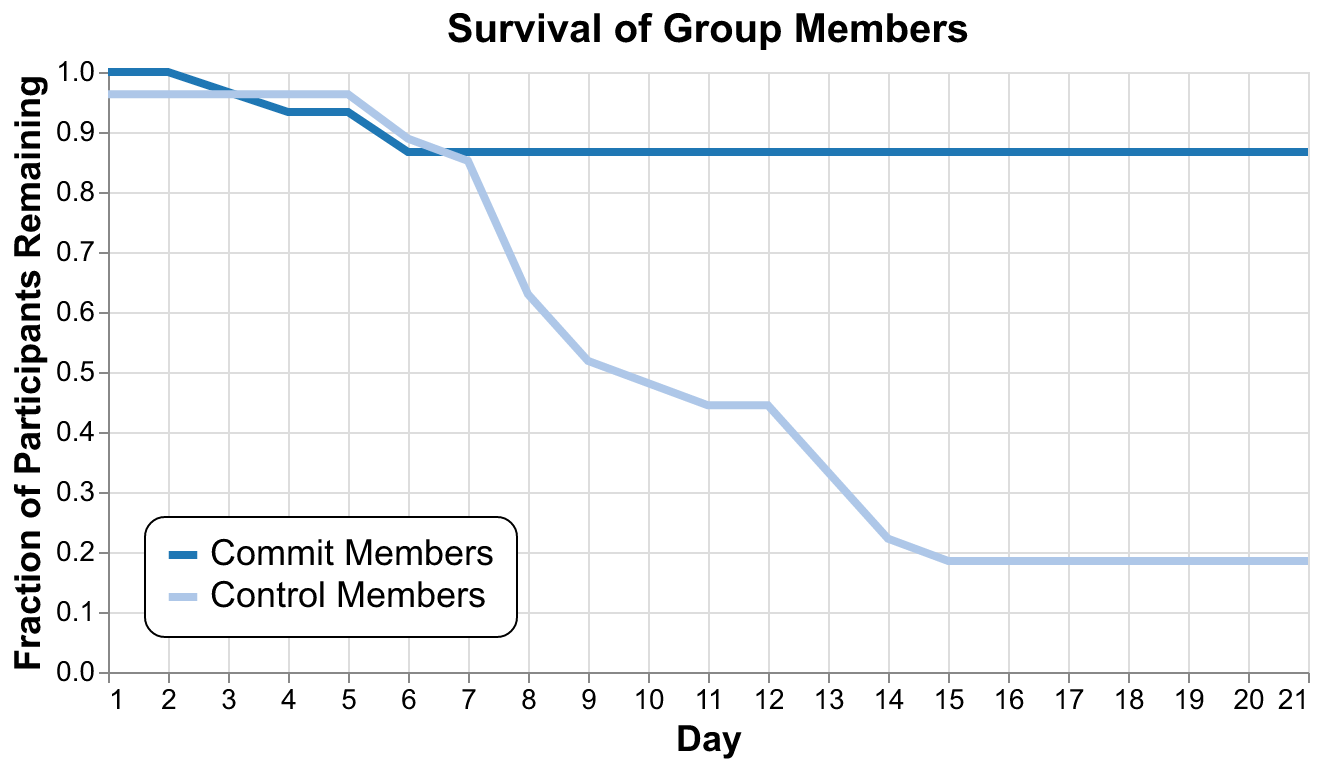}
  \caption{The number of surviving active participants across the study period. Participants ``survive'' so long as they are not inactive a for full week; if they are inactive for that length of time, they are no longer surviving, even if they return to post after that. By the end of the study, Commit condition groups had over four times the number of ``surviving'' active members. }
  \label{fig: participant survival}
\end{figure}

Groups in the Commit groups lasted longer, supporting Hypothesis 2 (Figure~\ref{fig: participant survival}). Far more participants in the Commit condition had stayed active until the end of the study than the control: 87\% versus 19\%, a factor of over four times more. This means that over six times the number of control participants spent an entire week inactive compared to those in the Commit condition.\footnote{Participants are defined as surviving so long as they have no been inactive for seven days in a row. They therefore cannot ``come back to life'' by renewing activity in the future. Other survival measures suffer from the question of whether participants might start being active after the study period concludes.} 

Conversation and activity did decrease some over time in each of the conditions, though significantly more so in the control condition.

We analyze the significance of the impact of the commitment mechanism on survival using a Cox proportional-hazards model (Table~\ref{table: survival model results} in Appendix~\ref{appendix: survival}). Condition is highly significant ($p<.001$) in predicting the longevity of participant activity in their group. Exponentiating the log hazard ratio, $-1.0137$, yields an estimate of less than half hazard ratio ($e^\beta=.363$) for the Commit condition. In other words, participants in the Commit condition had under half the odds of remaining inactive for a full week compared to those in the control condition, compounding each day. These activity survival results are robust to different conditions of ``surviving'': the same significant difference in longevity holds for analysis conducted for 3-9 days (Table~\ref{table: survival model results}).  

\subsection{Commitment mechanisms changed the group environment in ways that made it more comfortable to contribute}

Our goal with this system was to change individual behavior in ways that also changed the overall group environment, which would then come full circle to impact the individual experience of membership and participation in the group. Towards evaluating this goal, we sought to learn from participants their perception of the group environment--in terms of motivation, comfort, psychological safety, and relationship development--as well as how these factors impacted their interactions in the group. 

Commitment changed participants' framing of participation: as one participant shared, ``I think I was motivated to contribute just because it was called `commitment' \ldots like I didn't think it was a burden, but at the same time thought it was my responsibility to contribute'' (P39). 
Participants shared that the notifications in both conditions drew them back to the app and prompted them to respond to ongoing conversations. They felt strongly about trying to reply to each other's messages, with only forgetfulness or inapplicability dissuading a response. The desire to reply could be out of generic social obligation, as was the case for P39 who explained, ``I just didn't really want anybody's prompt to go unanswered, because that wouldn't feel very nice.'' The idea of commitment also reinforced this obligation for some participants:
\begin{quote}
    \textit{I don't know. I think I felt bad about not fulfilling the commitment. So like; it wasn't like---I ignore notifications all the time right, but because each person, I knew that each person in the group also had to like be committed to it. I felt I would feel like a little guilty if I didn't do my part to help keep the conversation going.} (P23)
\end{quote}

Social obligation was internalized for participants, rather than imposed upon them by other group members. That is: group members did not report policing each others' commitments or participation, but instead self-enforced out of a sense of duty. However, participants did report being aware of the participation levels of their group members, and in particular, noticing when certain members were rarely or always active.

Most everyone was willing to respond, but the need to initiate conversation made members self-conscious and unsure. Participants initially framed this concern around ``not knowing what to say,'' but when prompted about their specific concerns, revealed worries about reception and social acceptance. 
\begin{quote}
    \textit{I don't know what people are gonna be like, you know, I guess. Heck, is this the group where we pretend that everything is okay?} (P17)
\end{quote}

In these cases, Commit served as a gentle impetus for starting new conversations when activity stalled. Participants in the Commit condition would take it upon themselves to come up with another topic or question. Sometimes they framed it as their turn to take responsibility, like P9 who shared ``I'm gonna send a message and I'm gonna pick a new topic because I want other people to also do their part kind of, you know, so I can make a good conversation happen.'' Other members were more intrinsically motivated to converse, as P21 shared, ``I was getting kind of nervous about the commitment ending, and nobody, like, contributing, and \ldots I really enjoyed, you know, talking with these people, and I was trying to think of things that they would be interested in.'' Fulfilling was also a good excuse to take the initiative to post, which made P47 feel less self-conscious:
\begin{quote}
    \textit{It's not as strong as a safety net, maybe, but something similar where it kinda gives you like an excuse \ldots So that it doesn't, you know, you're not worried about people not responding because you're just supposed to do it, anyway.} (P47)
\end{quote}
Since the commitment reminders went out to all group members at the same time, it drew group members together, with P2 noting that they were often around to reply immediately to new conversations for that reason: ``I think we all kind of got [the notifications], or it seemed like we all kind of got them around the same time, and so somebody would send something.''

When participants were particularly unsure about what would resonate with their groups, they would resort to inoffensive questions---not necessarily creative or exciting, but easy to send and reply to. 

Sometimes, these innocuous prompts took off. For example, in one group, a member shared about the cat they were pet-sitting because, ``The possibility of people not liking pictures of cats was very low, as we, as we can probably gather from the Internet'' (P40), which became a much longer discussion as other participants also shared their pets. This moment ended up being a turning point for the group:
\begin{quote}
    \textit{The comfort level did increase \ldots people were just discussing or sharing information about their own cats and everything \ldots After that point, people were definitely more comfortable. They were sharing their plans for going home.} (P39)
\end{quote}

The impact of this small change is even clearer when we examine the data at the the time scale that participants were encouraged to participate at: two day periods. Both conditions encouraged participants to contribute to their groups at least once every two days, but only in the Commit condition did participants regularly meet this minimum. Participants in the Commit condition sent a median of one message per two-day period, while those in the control sent a median of zero. Between encouraging conversation initiation and swaying participants to reply more, commitment is very effective at drawing participants up to the level of expected participation that we set.

\subsection{Participants in Commit condition felt safer and more able to rely on each other}

For our participants, the impact of commitment was not only realized through increased messages sent, but through differences in the atmosphere and experience of the groups. We hoped to identify how these affective differences may have impacted the behavioral outcomes and vice versa. Participants in the Commit condition rated significantly ($t(46.8) = 2.55, p = 0.014$) higher in questions related to psychological safety in our survey (Table~\ref{table: survey response} in Appendix~\ref{appendix: survey}), and also reported back more comfortable group environments in the interviews.

Across both conditions, many participants described their groups as friendly but removed:
These groups still fell on a spectrum of activity levels and conversation quality, but the majority of groups did not form intimate connections with strangers during three weeks. Their members would describe the conversations as mildly interesting but disconnected:
\begin{quote}
    \textit{Almost like a networking night, except not really focused about any particular business topics, just like kind of like people who don't really know each other, but are kinda just like having friendly banter, and just like talking about things that one or a couple might be interested in.} (P10)
\end{quote}
However, several groups that were more active saw the atmosphere change with time. One participant remarked, ``I think just in conversing with others, it humanizes them more. So you feel like you have a little bit of an obligation to keep conversations going and to engage and dialogue with others'' (P49). Similarly, P45, whose group celebrated one member's birthday during the study, said, ``I think people did become more comfortable as time went on \ldots maybe I did see that like people did share more of personal details or personal anecdotes \ldots later on in the study.'' Through extended conversational efforts, several groups in the Commit condition ended up having unusually lively and involved conversations that led to a more friendly and informal environment. 
Interviews suggested that Commit groups reached this outcome because they had a higher incidence of strong, memorable conversations that drew members together.
The reason for this appears to be that participants in the Commit condition started more conversations, and therefore had more ``shots on goal'' increasing the likelihood that one of the conversations would really spark.

While every group had at least one conversation that did not take off, these did not have a strong impact on the group's trajectory.
One really good conversation, on the other hand, could be a turning point for the group. The cat conversation increased the comfort and informality of one group significantly. After the study, two members of that group shared that the group ``could have led to a good friendship'' (P38). More than one group became comfortable via extended discussions: a different one covered many potentially controversial topics including student unionization and visa processes. Members of this group described very positive experiences, and even planned to meet up in person.  
Commit groups did not lack for shallow, water-cooler level questions, but were more likely to also have ones that touched on complex or deeper topics.

Commit participants also described the effort as more shared. 

The commitment feature helped them feel a communal effort in continuing the conversations. P19 shared that ``I could feel that we all wanted to contribute to the conversation like because we kept like committing to participate in the cycle using the app.'' Other members believed their group adopted a team-like mentality in keeping the discussion going, with P21 saying ``I feel like this is like a group responsibility. So it's not like on anyone's particular shoulders.''

\begin{figure}[htb]
  \centering
  \begin{subfigure}{0.4\textwidth}
    \centering
    \includegraphics[width=\textwidth]{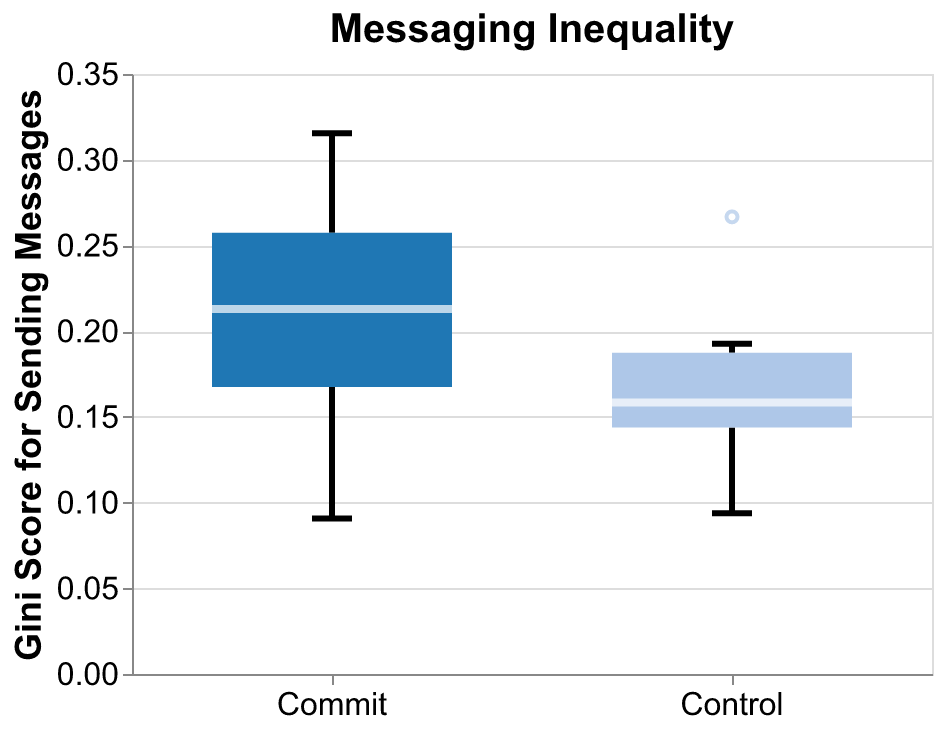}
  \end{subfigure}
  \hspace{0.04\textwidth}
  \begin{subfigure}{0.4\textwidth}
    \centering
        \includegraphics[width=\textwidth]{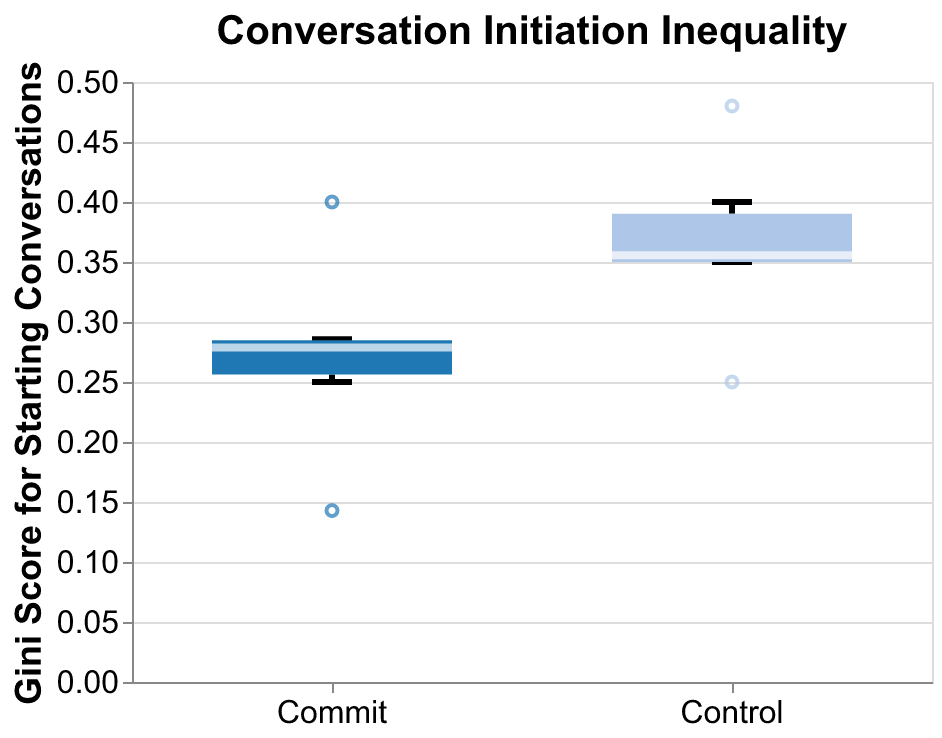}
  \end{subfigure}
  \caption{Comparisons of the equality of member contribution within groups for (a) all messages, and (b) messages initiating new conversations, each of which are measured across the entire study period. Inequality is measured using the Gini coefficient, for which higher values indicate greater inequality. Neither difference is significant, but we can see the directionality of the difference switch. }
  \label{fig: inequality boxplots}
\end{figure}

Participants in the Commit condition rated significantly ($t(50.4) = 2.28, p = 0.027$) higher levels of group equality in our survey, in addition to describing shared effort in the interviews. If we define effort simply in terms of messages sent, this is not the case. The median Gini coefficient,\footnote{The Gini coefficient is a measure of statistical dispersion, designed to reflect income inequality but transferable to other contexts. A Gini coefficient of 0 indicates perfect equality, and of 1 indicates maximal inequality. In our setting, 0 would mean that each participant had messaged the same number of times, and 1 would indicate that only one member had ever messaged.} a measure of inequality, for the message distribution across participants in a group is 0.21 in the Commit condition and 0.16 in the control, indicating greater inequality in Commit (though further analysis shows these results to not be significant, Table~\ref{table: gini results} in Appendix~\ref{appendix: inequality}). However, participants described an difference in effort for starting versus continuing conversation. So, when examining starting conversations---we operationalized ``starting'' as sending a message at least 12 hours after the last message sent---the median Gini coefficient is 0.28 for Commit versus 0.36 for control, flipping the directionality of difference (though further analysis again shows only marginal significance). Given that reciprocating effort was a major focus for Commit participants, it seems like the members may have shared the \textit{effort}, rather than simply sending the same amount of messages.

This equality result helps to disentangle the myriad impacts of Commit. While the majority of Commit condition participants described motivation coming from both the mechanism itself and the activity of their group members, those two effects are difficult to separate. However, from our data suggesting increased equality of conversation-starting, we have quantitative evidence of the more ``direct effect'' in that a greater share of Commit group participants were taking solo initiative. 

\subsection{Commitment is not a panacea for non-viable groups}

While Commit had many positive outcomes in our study group, the mechanism is insufficient to overcome cases when the group members are not invested in the group. Commitment is a tool for helping \textit{enhance} motivation in groups, not creating it entirely. We focus in this section on the subset of participants in the Commit condition who expressed unhappiness with their group outcomes. They reported complaints with the commitment mechanism itself, as well as about behavior in their groups that commitment failed to assuage.

\subsubsection{Notifications were unpleasant when members did not want to engage}

Some participants did not have an intrinsic desire to participate, and therefore resented the frequent notifications to send messages. The notifications did not have a ``delay'' option if people wanted to take a break; they would have to wait for the notifications to stop and then remember to come back themselves. Some participants who were busy or were having unusually difficult times offline found these irritating, with one sharing that, ``I think around Thanksgiving break was when I found the reminders or notifications to be little bit more troublesome, because I was engaged in other things'' (P40). Similarly, some participants thought the notifications were too frequent, and therefore would not always commit or fulfill their commitment. This finding underscores the importance of setting the correct commitment cycle length, because it will turn away group members if it is too demanding.

\subsubsection{Commitment sometimes drew out shallow conversation for longer}

We found that more conversation benefits groups by increasing the chances of stumbling onto an engaging topic of conversation. However, some participants instead experienced prolonged, shallow conversations. P49 shared that, ``The conversations never felt like they were going in-depth about something. It felt like I was answering the question for the sake of answering the question.'' 

Some participants also indicated that their involvement in the discussions was driven more by a sense of obligation than by enjoyment, merely fulfilling their commitment to keep the conversation going:
\begin{quote}
    \textit{I think, from my perspective, the conversation wasn't very interesting that I like, yeah. I think that we were trying to to maintain the conversation alive. But I think, at least for me, it never got very interesting.} (P19) 
\end{quote}
Overall, although commitment could sustain conversation, this may not be a desirable goal for groups who cannot find a topic of shared interest. Ultimately, participants who just want to lurk, or are not intrinsically motivated to participate, may find commitment stressful or annoying. Outside of a compensated study context, these members would likely drop out of the group.

\section{Discussion} \label{sec: discussion}

Through a three week field experiment, we observed that a commitment-based design encourages members to contribute more than a default reminder and encouragement scheme. The greater participation corresponded with greater group longevity and experiences of increased trust and comfort. 

Commitment is one part of a larger vision of reimagining the possible designs of social media platforms and how they support groups and communities. While Commit sought to rethink the design of a group chat, and is focused on facilitating the survival of small groups, there are remaining groups that are not well-supported by the existing design paradigms online.  
In this section, we discuss the broader set of opportunities in designing online spaces to meet the needs of groups and communities. We then examine in greater depth the social mechanism of commitment's success and how this may generalize to other domains and purposes. We conclude by reflecting on the limitations of our approach and evaluation design, with recommendations for improvements.

\subsection{Designing Away From Individualism and Expansion Online}
The modus operandi of most major social media platforms is to optimize toward growing and increasing the engagement of their user base. This is no secret: countless features on these platforms---such as the feed algorithms, the notifications and associated red badges, the friend or group recommendations, and suggestions to invite contacts---seek to bring more users on the platform and get them to come back again and again~\cite{narayanan2023understanding, backstrom2016billion}. 

However, this growth objective has side effects, even beyond the poor long-term individual impacts like social media addiction~\cite{hou2019addiction}, and the longer-term societal impacts like political polarization~\cite{kubin2021role}. A focus on drawing in and pleasing individual users also erodes the social contract of online interactions: there is no guaranteed shared context, no enforceable social obligation, and too many options to invest in any one, a common problem in online contexts~\cite{dangelo2017fish, moser2017choice}.

Commitment as a design lever reminds us that not every group or context demands high scale~\cite{hwang2021why}, and many online spaces would benefit from reduced size and heightened intimacy. Scale often comes at the cost of intimacy, comfort, quality, and strength of purpose. Our study participants specifically mentioned the small size of their group chats as a reason for participating, noting that the size of other groups they participated in (e.g., Discord servers) as a major reason behind their lurker status there. In the case of P15, it was not just the size of the group, but also the lack of specific incitements to contribute:
\begin{quote}
    \textit{I think it was actually less comfortable than this one, because like, I feel like the discord server is like, first of all, bigger. So I feel even less comfortable talking. And also there's like no obligation to send any messages, whereas, like for this one, since, like the app reminds me, it'd be like, `Oh, like, I can. I should probably like check in and like, see how everyone's doing.' Like I felt more motivated to like keep tabs on what everyone was doing, and it did feel a lot more comfortable.} (P19)
\end{quote}

This implication for design might feel at odds with traditional decision-making strategies on platforms. First, we would argue that higher-quality social interaction should trump higher volumes of social interaction. But further, we argue that the two are not necessarily even in tension: as Tom Cunningham at the Integrity Institute, and formerly of Facebook and Twitter describes, ``Platforms care primarily about long-run retention, engagement is a means to that end [...] They also would sacrifice substantial short-term DAU [daily active users] if it could be shown with confidence that it would lead to higher long-term DAU.'' In other words, if commitment leads to fewer members in each group, but members remain more committed to their groups over longer time periods, many platforms might agree to that tradeoff.

\subsection{The Power of Intervening on Groups, Together}

While a major piece of commitment's efficacy was making membership effortful, it also took advantage of the multiplicative impact of interventions made across an entire group. Had commitment been an intervention on an individual, where only one group member was aware of it, it would have been far less effective. Many participants in the Commit condition noted that they trusted other members to respond, and that the responsiveness they had seen was a factor in their continued participation efforts. 

Opening up and self-disclosing created a similar feedback loop; P20 shared that, ``Over time, people were more willing to open up and I think that's because there were some folks who really opened up [earlier on].'' In the design of Commit, we made use of the knowledge that group behavior is linked~\cite{burke_feedme} to assuage participant worries about reciprocation. 

The discussion starters found reassurance in the knowledge that their group members needed to fulfill their own commitments---this increased the chances that someone would reply to their conversation prompt rather than leaving them hanging. P47 explained, ``So then, even though I know they might not fulfill it because of life reasons \ldots it does increase the probability that they will, they will reply, and that's also helpful.'' They went on to explain that the Commit reminder alone did not remove the discomfort of no one replying, without the reassurance of the group being reliable:
\begin{quote}
    \textit{I would still feel bad if I post something because of a commitment reminder and nobody responds, I would definitely still feel bad, even though posting would be made easier because I need to fulfill commitment. So knowing that other people would likely respond because of their commitment requirement too also makes it easier to to post a new message, and then also makes the waiting a little easier. Because you're just expecting someone to respond, anyway.} (P47)
\end{quote}

We believe commitment worked because the small changes it incited in a single member were visible to the other members~\cite{erikson2000translucence} and improved their experience, therefore influencing their behavior as well in a positive feedback loop of increased participation. Figure~\ref{fig: commit mechanism} describes this proposed mechanism that we synthesized from our interviews and data.

Commitment's success therefore offers insight around increasing the efficacy of other interventions to do with groups and social activities. In cases with similar critical mass issues, where people do not want to take action without guarantee of benefits or the action is more difficult to make when made alone, some sort of visible pledge~\cite{cheng2014catalyst} may increase their willingness to act.

\subsection{Commitment Design Space} \label{design space}

We instantiated the Commit app around the setting of new group chats. Our setting also involves a flat hierarchy, and social motivations rather than any sort of external purpose around production of goods or software. These factors led to different decisions in the design of Commit. 

Since the groups are small, we did not allow for a ``null commit'' option, in which members promise no level of participation and still gain entrance to the group. The participation of each member is so critical in small groups, so we did not want to normalize inactivity. We also set the commitment cycle to a short period of time, two days, to keep a consistent level of contribution. Since these groups are both small and socially oriented, we did not include a punishment or enforcement mechanism behind failing to fulfill a commitment. We judged that social pressure is sufficient for encouraging people to live up to their commitments.

Finally, since this app is designed for those who are starting a social group (rather than sustaining one), there is no mechanism to change the commitment process over time or scaffold users into different roles as they become more senior.

However, we believe that there are good reasons to change the design of commit, and we review examples of relevant contexts below. 
\begin{itemize}
    \item In a group with a \textit{notable hierarchy structure} (e.g., a workplace social channel), there could be \textit{different commitment levels} that different levels of the hierarchy choose (perhaps the leader commits to starting discussion versus the members who commit to replying).
    \item In a context in which \textit{losing members is acceptable or unused spots have a notable cost} (e.g., an exclusive feedback group for artists), \textit{stronger commitment enforcement could be set}, like forfeiting one's spot after continued failure to fulfill a commitment.
    \item In contrast, for a group in which \textit{more members are mostly beneficial}, even if they are inactive, \textit{a null (no required action) commitment or reaction emoji commitment could be added}. In this case, simply asking members for a level of commitment may nudge many to engage, even if the option is present for them to avoid the commitment---they have to select it. 
    \item In a group where \textit{conversation is more sporadic} or less synchronous due to the context (e.g., a book club that is reading one book each month), \textit{the commitment cycle can be lengthened} to allow for asynchronous discussion at each member's pace.
\end{itemize}

Setting the contextualized design of the commitment system takes care, and poor decisions may have significant repercussions for its efficacy. For example, setting too low of a commitment expectation may prevent commitment from having any effect. However, too high of a threshold may dissuade potential members from even joining the group. In general, we recommend setting it to a single unit of participation (one message, for example) per unit of time that you wish for there to be significant group interaction. Since we wanted close to sustained interaction (multiple conversations per week), we chose every other day. However, as in our book club example, if contribution is only desired on a monthly cadence, then a month is a sensible commitment period.

Even beyond these parameter level changes, additional tweaks can be made to the design to better serve a group's current state. For example, if members are becoming disenchanted with the repeated action of commitment and intend to be active anyway, an auto-renewal could be instantiated that requires manual action only if they ever lapse on their commitment. A strong sense of commitment is only needed when members would not otherwise be active participants, so it is acceptable for commitment to fade into the background as the group strengthens. Additionally, if the group reaches a level of activity that appears to be self-sustaining, the group can try phasing out commitment. We recommend checking in with the group members regularly, and at any sense of discontent with the mechanism, to discuss how to modify it to meet the group's current needs.

\subsection{Limitations}
While the results of our evaluation are promising, we cannot claim that they will generalize to different group contexts online. We recommend adaptions for the commitment mechanism to different contexts, but these are limited by what we have been able to observe. 

We engineered a socially difficult scenario to test commitment. For many of the participants, being matched in groups with strangers without any guaranteed shared background or scaffolds was stressful. Our goal was to see the extent of what Commit was capable of overcoming. However, this does produce the caveat that the difference across conditions could be less extreme in a more comfortable setting. Beyond being socially uncomfortable, the study scenario is also distinct from most online groups where membership is intrinsically motivated, often by a common interest or existing social ties. While we matched participants based on loose common interests, they likely had lower intrinsic motivation to join and participate in their groups than in a standard online group, and instead were extrinsically motivated by the study context. We therefore cannot be completely sure about the impact of commitment in  a true online context. However, it has shown promising results in two different deployments within our research community.

We also set a strong control in order to challenge the advantage commitment poses versus other existing approaches. Our control condition was not just the absence of the commitment feature, but an alternative that sent reminders and urges to action at the same cadence, and included similar banners that alerted users to their contribution frequency.

This design had a positive impact in terms of validity: participants in the control condition largely thought they were in a treatment condition due to receiving notifications that seemed like interventions, which helps prevent participant bias from demand characteristics from skewing our comparison.

Commit may not be successful or necessary in all contexts, for example, groups much larger than the four or five members of our evaluation. In much larger groups, the number of members makes it harder to observe who does or does not fulfill their commitment, and therefore may make the mechanism less effective. Large groups may also already have sufficient activity due to size. Commitment does not make sense either in significantly smaller groups, in which social loafing cannot really exist (e.g., a two-person conversation). It could be that commitment loses its efficacy over time, given that our study only lasted for three weeks, though our non-study deployments have continued using the mechanism for many months. 

We also do not know how well commitment would work outside of our study population, which largely consisted of college students who were frequent group chat participants. In countries outside of the United States, there may also be different cultural norms at play which would affect the efficacy of commitment. For example, the cultural dimension of individualism versus collectivism could impact the need to explicitly direct group members towards thinking about collective needs. 

\subsubsection{Drawbacks of Commitment}

Commitment is an intervention for certain groups, and has its inherent trade-offs. We want to highlight how enforcing commitment could actively cause harm. Commitment explicitly trades away individual autonomy for group benefit. Individuals may feel stressed about the commitment, due to busyness or social anxiety. Commitment should therefore be deployed in contexts where the members or leadership feel like the benefits of increased participation are worth the negatives of increased pressure. Groups may also make choices to help out newcomers or anxious members like creating a grace period when joining or allowing for null commitments. We also recommend avoiding structures that ``name and shame'' those who lapse or fail to fulfill their commitments. 

In cases where there is a real concern for harm to group members from urging them to participate, we recommend reducing the pressure to participate and the visibility of group members' participation. For example, in a group with frequently combative political discussions, a group could create a null commitment option that not only stopped encouraging contribution, but also reduced members' visibility (by not showing them among group membership, for instance). This option could be taken by group members who are marginalized with respect to the group topics, if there is fear of harassment or tokenization of those members. 

Even with these considerations, commitment may be too strong of a push for some members, similar to how participation requirements in classes alienate some students. This is a trade-off that our design makes, to support groups that cannot otherwise thrive by asking more of each participant. However, our intention is not to mandate commitment across all online groups, but rather to suggest their ability to productively add to the online ecosystem, like how seminar classes complement lecture-style classes that do not require participation without making them obsolete. Also, while online groups have long facilitated no-commitment membership, most offline social spaces do not: members of a club who never attend meetings are not, in practice, \textit{in} the club. Commitment facilitates lively, participatory online spaces to grow and survive in spite of the dominant inclination to lurk online.

\section{Conclusion} \label{sec: conclusion}

In this paper, we introduce \textit{commitment} as a design strategy for online groups, and instantiate it in the  Commit system. Commitment changes the nature of online group membership: what it asks of us, what it signals to others, and what we receive back. By setting and communicating higher expectations of group members, the commitment mechanism encourages members to invest effort in the group and expect the same of their compatriots.
In doing so, commitment supports small groups to encourage the level of participation they need to survive. A three week between-subject field study demonstrates that commitment elicits the desired effects: increased contributions from users, longer group survival, and stronger feelings of comfort and support in the groups. As informed by our evaluation and design process, we provide recommendations to adapt the commitment mechanism to different group contexts and discuss the greater design space of designing online environments to support group and communal interaction.

\begin{acks}
We thank our anonymous reviewers, as well as Omar Shaikh, Joon Sung Park, and Catherine Han for their feedback on the paper. Additional thanks go to Helena Vasconcelos, Jordan Troutman, Michelle Lam, and Tiziano Piccardi for their assistance in early prototyping. This work was supported in part by NSF grant CCF-1918940 and Apple.
\end{acks}
\bibliographystyle{ACM-Reference-Format}
\bibliography{main}
\appendix
\section{Appendix}

\subsection{App Notifications} \label{appendix: notifications}
\FloatBarrier

\setlength{\tabcolsep}{6pt}
\renewcommand{\arraystretch}{1.5}

\begin{table}[htb]
\small
\begin{center}
\begin{tabular}{p{2.5in} p{2.5in}}
\hline
Commitment Condition & Control Condition \\
\hline
\textit{At the end of the two-day commit cycle, if they do not re-commit}: \newline
Your commitment to [group name] has lapsed! Make sure to come back and re-commit so you can continue seeing content. & 
\textit{After two days without checking group}: 
\newline
You haven’t checked [group name] in several days! Come back and check out what you've missed.
\\
\textit{After they have been lapsed for a full two-day commitment cycle}:
\newline
Your commitment to [group name] has been lapsed for a cycle. Do you want to recommit?
&
\textit{After four days without checking group}:
\newline
It’s been a while since you've visited [group name]! Come back and check out what you've missed.
\\
\textit{The morning of a new cycle when commitment has lapsed}:
\newline
A new commitment cycle is starting! Make sure to come back and re-commit so you can continue seeing content in [group name] has lapsed! 
&
no equivalent notification
\\
\textit{At the end of the two-day commit cycle if they have not participated}:
\newline
The commitment period for [group name] is close to ending and you have not contributed yet. Come back and share your thoughts!
&
\textit{After two days without messaging}:
\newline
You haven’t messaged in [group name] since several days ago. Come back and catch up!
\\
no equivalent notification
&
\textit{After four days without messaging}:
\newline
You haven’t messaged in [group name] since several days ago. Come back and catch up!
\\
\end{tabular}
\end{center}
\medskip
\caption{Notifications were paired, so that each had an equivalent in the other condition that was sent with the same frequency and for a similar reason. We were unable to match two (one in each condition).}
\label{table: notifications}
\end{table}

\FloatBarrier

\subsection{Survey Questions} \label{appendix: survey}

We administered short surveys to every participant at the end of the study period. Each question was answered according to a seven-point Likert scale, where responses ranged from ``Strongly Disagree'' to ``Strongly Agree.'' Use of * indicates that the survey item was reverse coded.

Valued and important
\begin{enumerate}
    \item My participation positively impacts the group.
    \item My contributions are valued by other members of the group.
    \item This group appreciates my differences and unique perspective.
\end{enumerate}

Safety
\begin{enumerate}
    \item I feel comfortable taking social risks in this group (e.g., reaching out first).
    \item I have been left hanging by other members of this group (e.g. I sent a message that no one replied to).*
    \item I have avoided sending a message out of worry it will not be received well.*
\end{enumerate}

Commitment
\begin{enumerate}
    \item I feel some responsibility for this group and the outcome of conversations here.
    \item I make a conscious effort to participate regularly in the group.
    \item I care about making this group succeed.
\end{enumerate}

Inequality
\begin{enumerate}
    \item I or another member carry this group (i.e., without their contribution, the group would be quiet/unsuccessful).*
    \item The same people (or person) always start the discussion.*
    \item Some people never reply or contribute to the discussion.*
\end{enumerate}

\subsection{Interview Themes} \label{appendix: interview}

We conducted semi-structured interviews with participants to uncover the subjective experiences of the participants in the different conditions of the group chat and understand possible mechanisms that explain our results. Our interviews adapted for each participant to dive deeper into relevant details and experiences they brought up. The questions below are not an exhaustive list of what we asked in the interviews, but reflect the approximate topics and themes we tried to cover.

\begin{itemize}
    \item What was the group atmosphere?
    \item What was the topic and quality of discussion?
    \item How comfortable and familiar did they feel with other members?
    \item What was the activity level, for the individual and the group as a whole?
    \item How did the activity change with time?
    \item How was participation divided across members? (starting discussions, replying, reacting)
    \item What was their motivation to participate? Did they feel obligated?
    \item Which factors, if any, discouraged them from participating?
    \item How did they perceive and respond to the notifications?
\end{itemize}

\subsection{Activity Analysis} \label{appendix: activity}

\FloatBarrier


\begin{table}[h] \centering 

\begin{tabular}{@{\extracolsep{5pt}}lcc} 
 & \multicolumn{1}{c}{\textit{Active Status}} & \multicolumn{1}{c}{\textit{Log(Messages Sent)}}\\ 
  & \multicolumn{1}{c}{Mixed effects logistic regression} & \multicolumn{1}{c}{Linear mixed effects model} \\
\hline
 Commit Condition & 1.257$^{***}$ \hspace{0.2cm} (0.186) & 1.143$^{*}$ \hspace{0.2cm} (0.467) \\
 Full Group & $-$0.431 \hspace{0.2cm} (0.227) & $-$0.793 \hspace{0.2cm} (0.543) \\ 
 Day & $-$0.071$^{***}$ \hspace{0.2cm} (0.011) & - \\ 
 Intercept & $-$0.302 \hspace{0.2cm} (0.189)  & 2.503$^{***}$ \hspace{0.2cm} (0.387) \\ 
\hline
\end{tabular} 
\medskip
\caption{\textit{Active status} (N=1,197): we observe an odds ratio of 3.5 (log odds of 1.257) for activity of one member on a given day based on condition, Commit versus Control. This effect is highly significant. \textit{Messages sent} (N=57): we observe an increase in the logged number of messages by 1.143 (an increase by roughly a factor of 3), a significant effect. Note that activity is measured on a day-by day-basis, while messages sent are analyzed across the entire study period.
Also note: $^{*}$p$<$0.05; $^{**}$p$<$0.01; $^{***}$p$<$0.001.}
\label{table: activity model results}

\end{table}

\FloatBarrier

\subsection{Survival Analysis} \label{appendix: survival}
\FloatBarrier

\begin{table}[h] \centering 
\begin{tabular}{@{\extracolsep{5pt}}lccccc} 
 & \multicolumn{5}{c}{\textit{Dependent variable: Survival}} \\ 
\cline{2-6} 
Lapse Period (Days) & 3 & 5 & \textbf{7} & 9 & 11 \\ 
\hline 
 Commit Condition & $-$1.513$^{***}$ & $-$1.407$^{***}$ & \textbf{$-$1.014$^{***}$} & $-$0.623$^{*}$ & $-$0.480 \\ 
  & (0.350) & (0.300) & \textbf{(0.279)} & (0.268) & (0.267) \\ 
\hline 
Observations & 57 & 57 & \textbf{57} & 57 & 57 \\ 
R$^{2}$ & 0.301 & 0.305 & \textbf{0.198} & 0.088 & 0.054 \\ 
Max. Possible R$^{2}$ & 0.998 & 0.998 & \textbf{0.998} & 0.998 & 0.998 \\ 
\hline 
\end{tabular} 
\medskip

\caption{Results of the Cox Proportional-Hazards Model for survival analysis. The log hazard ratio is $-1.0137$. The Commit condition therefore has less than half the hazard ratio ($e^\beta=.36$). We interpret this to mean that participants in the Commit condition had under half the odds of being inactive for a week compared to those in the control condition, compounded each day. Note: $^{*}$p$<$0.05; $^{**}$p$<$0.01; $^{***}$p$<$0.001.}
\label{table: survival model results}
\end{table} 
\FloatBarrier
\subsection{Survey Measure Analysis}
\FloatBarrier
\setlength{\tabcolsep}{6pt}
\renewcommand{\arraystretch}{1}
\begin{table}[htb]
\begin{center}
\begin{tabular}{l@{\hskip 0.5in} c c c} \\
Theme & t-statistic & p-value & df \\
\hline \\[-2ex]
Safe & 2.55 & 0.014 & 46.8 \\
Valued & 1.41 & 0.17 & 48.1 \\
Equal & 2.28 & 0.027 & 48.3 \\
Committed & 0.902 & 0.37 & 50.4 \\

\end{tabular}
\end{center}
\medskip
\caption{Survey results, collected from participants after the conclusion of the group chat portion of the study. We analyze average response scores for each theme category with an unpaired t-test. (Response scores are calculated by assigning -3: Strongly Disagree, -2: Disagree, \ldots 3: Strongly Agree. Questions marked as reverse coded in Appendix~\ref{appendix: survey} are transformed by multiplying by -1, scores for disagreement and agreement switch.)}
\label{table: survey response}
\end{table}
\subsection{Conversation Inequality Analysis} \label{appendix: inequality}
\FloatBarrier
\begin{table}[h] \centering 
\begin{tabular}{@{\extracolsep{5pt}}lcc}  
\hline
& \multicolumn{2}{c}{\textit{Gini Inequality Coefficient for}:} \\ 
& All messages & Conversation-starting messages
\\ 
\hline 
 Commit & 0.041 & $-$0.093 \\ 
  & (0.040) & (0.045) \\ 
 Intercept & 0.169$^{***}$ & 0.365$^{***}$ \\ 
  & (0.029) & (0.032) \\ 
\hline
\end{tabular} 
\medskip
\caption{Comparison of Gini Inequality Coefficient for all messages and for starting conversations. The difference in inequality was not significant for all messages or just conversation-starting ones, though condition was marginally significant (p=0.069) for starting conversations. We note that the directionality of the difference does change when subsetting the messages to just look at those starting a new conversation. We can see a mild negative difference (greater equality) associated with the Commit condition when looking at conversation initiation, versus a mild positive difference (greater inequality) when looking at all messages sent. Note: $^{*}$p$<$0.05; $^{**}$p$<$0.01; $^{***}$p$<$0.001.}
\label{table: gini results}
\end{table} 
 \FloatBarrier

\end{document}